\documentstyle[prl,aps,float,multicol,epsf]{revtex}

\floatstyle{plain}  

\newfloat{FIG1}{ht}{lop}

\newfloat{FIG3}{t}{lop}

\begin{document}

\draft

\title{Detection of Coulomb Charging around an Antidot}

\author{M.~Kataoka,$^{1}$ C.~J.~B.~Ford,$^{1}$
G.~Faini,$^{2}$ D.~Mailly,$^{2}$ M.~Y.~Simmons,$^{1}$ D.~R.~Mace,$^{1,\ast}$
C.-T.~Liang,$^{1}$ and D.~A.~Ritchie$^{1}$}

\address{ $^{1}$Cavendish Laboratory, Madingley Road, Cambridge CB3 0HE,
United Kingdom }

\address{
$^{2}$Laboratoire de Microstructures et de Microelectronique - CNRS,
196,~Avenue Henri Ravera, 92220~Bagneux, France }

\date{\today}

\maketitle

\widetext 

\begin{abstract} 

\leftskip 54.8pt \rightskip 54.8pt

We have detected oscillations of the charge around a potential hill (antidot)
in a two-dimensional electron gas as a function of a perpendicular magnetic
field $B$. The field confines electrons around the antidot in closed orbits,
the areas of which are quantised through the Aharonov-Bohm effect. Increasing
$B$ reduces each state's area, pushing electrons closer to the centre, until
enough charge builds up for an electron to tunnel out. This is a new form of
the Coulomb blockade seen in electrostatically confined dots. We have also
studied $h/2e$ oscillations and found evidence for coupling of
opposite spin states of the lowest Landau level.


\end{abstract} 

\begin{multicols}{2} 

\narrowtext

Coulomb blockade (CB) in an open system sounds paradoxical. CB arises from
the discrete charge of an electron. For charging to happen, it has been
generally believed that electrons must be electrostatically confined in a
small cavity. Although it has recently been reported that ``open'' dots can
also show charging effects \cite{openCB,Marcus,CTLopenCB}, they are not
completely open systems, still having some degree of electrostatic
confinement.

In contrast, an antidot, which is a potential hill in a two-dimensional
electron gas (2DEG), is in a completely open system. Thus it has often been
assumed that CB does not occur when an electron tunnels through a state bound
around an antidot by a large perpendicular magnetic field $B$ ($> 0.2$~T).
Here, electron waves travel phase-coherently around the antidot with
quantised orbits, each enclosing an integer number of magnetic flux quanta
$h/e$ through the Aharonov-Bohm (AB) effect. Where the potential is sloping,
these single-particle (SP) states have distinct energies.  Conductance
oscillations observed as a function of $B$ or gate voltage have been
attributed to resonant tunnelling through such discrete states from one edge
of the sample to the other. This causes resonant backscattering or
transmission depending on the tunnelling direction \cite{Mace}. Up until
now, no charging effect has been taken into account in the system
\cite{Goldman,Maasilta}. However, Ford {\em et al.\/} \cite{Chris} proposed
that antidot charging should be present to explain double-frequency AB
oscillations, where two sets of resonances through the two spin states of the
lowest Landau level (LL) were found to lock exactly in antiphase, giving
$h/2e$ periodicity, and to have the same amplitudes in spite of different
tunnelling probabilities \cite{Chris,Sachrajda1}. There is as yet no full
explanation for these phenomena.

\end{multicols}

\widetext

\setlength{\textfloatsep}{1mm}

\begin{FIG1}

% FIG. 1.     

\begin{figure}[!ht]       

\epsfxsize=0.61\textwidth		   

\centerline{\epsffile{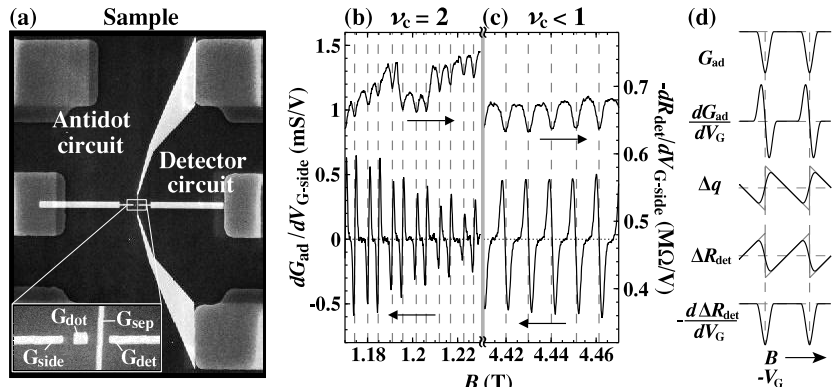}}  

\caption{(a) SEM micrograph of a device prior to second metallisation.   (b)
and (c) $dG_{\rm{ad}}/dV_{\rm{G-side}}$ of the antidot circuit and
$-dR_{\rm{det}}/dV_{\rm{G-side}}$ of the detector circuit with the gate
voltage on G$_{\rm{side}}$ modulated in two different regimes: (b) $\nu
_{\rm{c}} = 2$ and (c) $\nu _{\rm{c}} < 1$. Vertical dashed lines show the
alignment of the dips in the detector signal with zeros in the
transconductance oscillations. (d) Illustration of the relation between
various lineshapes. Grey lines in $\Delta q$ and $\Delta R_{\rm{det}}$ are
the ideal case, and black curves represent broadened lineshapes. }      

\label{fig:Det}         

\end{figure} 

\end{FIG1}

\begin{multicols}{2} 

\narrowtext

The aim of this paper is to demonstrate that magnetic confinement causes
charging in antidot systems, although there is no electrostatic confinement.
We have conducted non-invasive detector experiments \cite{Field} and obtained
clear evidence of charge oscillations around an antidot as a function of $B$
\cite{Kataoka}. We have also investigated $h/2e$ AB conductance oscillations.
The data show that the resonance only occurs through states of one spin,
explaining the matched amplitudes.

The samples were fabricated from a GaAs/AlGaAs heterostructure containing a
2DEG of sheet carrier density $2.2 \times 10^{15}$~m$^{-2}$ with mobility
370~m$^{2}$/Vs. An SEM micrograph of a device is shown in
Fig.~\ref{fig:Det}(a). A square dot gate (G$_{\rm{dot}}$), 0.3~$\mu$m on a
side, was contacted by a second metal layer evaporated on top of an insulator
(not shown) to allow independent control of gate voltages. The lithographic
widths of the antidot and detector constrictions were 0.45 and 0.3~$\mu$m,
respectively. All constrictions showed good 1D ballistic quantisation at $B =
0$. A voltage of $-4.5$~V on the separation gate (G$_{\rm{sep}}$), of width
0.1~$\mu$m, divided the 2DEG into separate antidot and detector circuits. The
detector gate (G$_{\rm{det}}$) squeezed the detector constriction to a high
resistance to make it very sensitive to nearby charge. To maximise the
sensitivity, transresistance measurements were made by modulating the
dot-gate voltage (or the voltage on the side-gate G$_{\rm{side}}$) at 10~Hz
with 0.5~mV rms and applying a 1~nA DC current through the detector
constriction. Simultaneously, the transconductance of the antidot circuit was
measured with a 10~$\mu$V DC source-drain bias, when necessary. The
experiments were performed at temperatures down to 50~mK.

Figures~\ref{fig:Det}(b) and (c) show the transresistance
$-dR_{\rm{det}}/dV_{\rm{G-side}}$ (transconductance
$dG_{\rm{ad}}/dV_{\rm{G-side}}$) vs $B$ of the detector (antidot) circuit in
two different field regions: (b) $\nu _{\rm{c}} = 2$ and (c) $\nu _{\rm{c}} <
1$, where $\nu _{\rm{c}}$ is the filling factor in both antidot
constrictions, which were determined from the conductance $G_{\rm{ad}}$. The
filling factors in the bulk 2DEG were $\nu_{\rm{b}}=7$ and 2, respectively.
The oscillations in $G_{\rm ad}$ occur as SP states around the antidot rise
up through the Fermi energy $E_{\rm F}$. The AB effect causes the overall
period $\Delta B$ to be $h/eS$, where $S$ is the area enclosed by the state
at $E_{\rm F}$.  The curve in (b) has pairs of spin-split peaks, whereas in
(c) only one spin of the lowest LL is present. The dips in
$-dR_{\rm{det}}/dV_{\rm{G-side}}$ correspond to a saw-tooth oscillation in
the change $\Delta R_{\rm{det}}$ from the background resistance (see
Fig.~1(d)). Here, note that a small increase in $B$ or decrease in
$V_{\rm{G}}$ has a similar effect on the SP states. Hence, integration with
respect to $B$ and $-V_{\rm{G}}$ are qualitatively equivalent.  Thus the net
charge $\Delta q$ nearby suddenly becomes more positive (making the effective
gate voltage less negative) whenever the antidot comes on to resonance (since
the dips line up with the zeros in $dG_{\rm{ad}}/dV_{\rm{G-side}}$). The
charging signals are not dependent on the presence of conductance
oscillations in the antidot circuit. It is still possible to observe the
signal with no applied bias in the antidot circuit, or when the side-gate
voltage is set to zero so that there is no tunnelling between that edge and
the antidot. Hence we conclude that this charge oscillation is associated
with states near the antidot, and interpret it as CB.

% FIG. 2.   

\begin{figure}    

\epsfxsize=85truemm  

\centerline{\epsffile{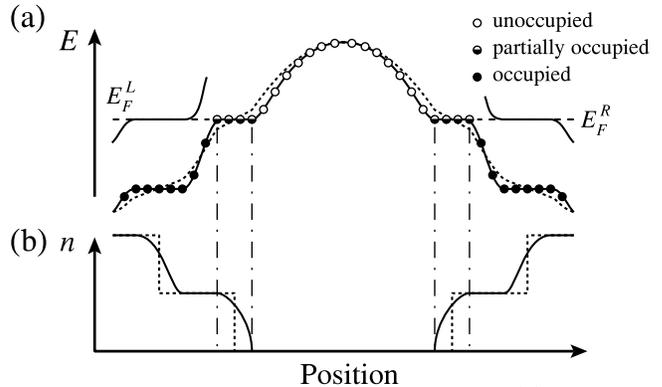}}         

\caption{Cross-section through the antidot: (a) energy of the lowest LL near
the antidot and (b) carrier density distribution. The conventional and
self-consistent pictures are shown as dotted and solid lines, respectively. A
bulk LL, which is reflected from the constrictions, is also shown.  The
vertical dash-dotted lines indicate the edges of a compressible region.}

\label{fig:Antidot}            
   
\end{figure}

Before showing how the charging occurs, it is worth reconsidering the shape
of the antidot potential. The conventional picture is a potential hill
smoothly increasing towards the centre as shown dotted in
Fig.~\ref{fig:Antidot}(a). However, for $B>0$, such a potential would require
abrupt changes in the carrier density where LL intersect $E_{\rm
F}$, which is not electrostatically favourable. Chklovskii {\em et al.\/}
\cite{Chklovskii} treated such a problem along the edge of a 2D system and
introduced alternating compressible and incompressible strips. Compressible
strips require flat regions in the self-consistent potential as depicted by a
solid line in the figure. It has always been considered that the potential
should not be completely flat in antidot systems \cite{Maasilta}, since the
presence of several SP states at $E_{\rm F}$ makes AB conductance
oscillations impossible in the simple non-interacting picture. However, if CB
of tunnelling into the compressible region occurs, conductance oscillations
with periodicity $h/e$ can still occur for such a self-consistent potential.

We explain the charging as follows.  As $B$ increases, each SP state
encircling the antidot moves inwards, reducing its area to keep the flux
enclosed constant. This results in a shift of the electron distribution
towards the antidot centre. One may think such a shift should not occur due
to screening in the compressible region. However, since each state is
discrete and is trapped around the antidot, and the incompressible regions
obviously have one electron per state, the total number of electrons in the
compressible regions must be an integer. Hence the compressible region also
moves inwards with the states. As a result, a net charge $\Delta q$ builds up
in the region. When it reaches $-e/2$, one electron can leave the region
and $\Delta q$ becomes $+e/2$. This is when resonance occurs, as for CB in
a dot.  At the same time, the compressible region, by losing the innermost
state and acquiring one at its outer edge, shifts back to its original
position just after the previous resonance. The same argument also applies,
of course, even if there is no compressible region, as the states are still
discrete.

As in quantum dot systems, the SP energy spacing $\Delta E_{\rm sp}$ and the
charging energy $e^{2}/C$ together determine when resonance occurs ($C$ is
the capacitance of the antidot). We have deduced these energy scales from the
temperature dependence of the charging signals and the antidot conductance
oscillations, and the DC-bias measurements of the differential antidot
conductance. The detailed analysis is given in Ref.~\cite{Kataoka}. We found
that $\Delta E_{\rm{sp}}$ decreases as $1/B$, as expected. In contrast,
Maasilta and Goldman \cite{Maasilta} found an almost constant energy gap,
which we interpret as the interplay of $\Delta E_{\rm{sp}}$ and a charging
energy which is small at low $B$ and saturates at high $B$.

The presence of charging should help to explain the $h/2e$ AB oscillations.
Fig.~\ref{fig:DBHIGH} shows AB conductance oscillations as both constrictions
are narrowed keeping the symmetry. On the $\nu _{\rm{c}}=1$ plateau the outer
spin state is excluded from the constrictions. Peaks up from this plateau for
$B<2.7$~T are due to inter-LL resonant transmission \cite{Mace}. This is only
noticeable when resonant backscattering is absent, i.e., on the plateau, and
is irrelevant in the arguments here. We focus on the resonant backscattering
process, which is caused by intra-LL scattering in the constrictions (see
diagrams at the right of Fig.~\ref{fig:DBHIGH}). The tunnelling probability
into the antidot states from the current-carrying edges is controlled by the
side-gate voltages.  The flat $\nu _{\rm{c}}=1$ plateau implies that there is
no tunnelling into the inner spin state. Hence, at higher $\nu _{\rm{c}}$
{\em at the same field}, where the constrictions are wider, there can also be
no such tunnelling, despite the presence of $h/2e$ oscillations. It is not
yet clear why the outer spin states should come on to resonance twice per
$h/e$ period; however, the equal amplitude of the resonances {\em can} be
explained since the tunnelling probability for that spin should be almost the
same for each resonance.

In conclusion, we have used a non-invasive charge detector to show
that tunnelling into antidot states is Coulomb blockaded. When states
of both spins are occupied, $h/2e$ oscillations are seen but
tunnelling is only via states of one spin, showing that there is a
strong coupling with states of the other spin.

This work was funded by the UK EPSRC. We thank C.~H.~W.~Barnes and
C.~G.~Smith for useful discussions. M.~K. acknowledges financial support from
Cambridge Overseas Trust. 

$^{\ast}$Present address: The Technology Partnership PLC, Melbourn Science
Park, Melbourn, SG8 6EE, UK.

\end{multicols}

\widetext

\setlength{\textfloatsep}{1mm}

\begin{FIG3}

% FIG. 3.     

\begin{figure}[!t]       

\epsfxsize=0.6\textwidth   

\centerline{\epsffile{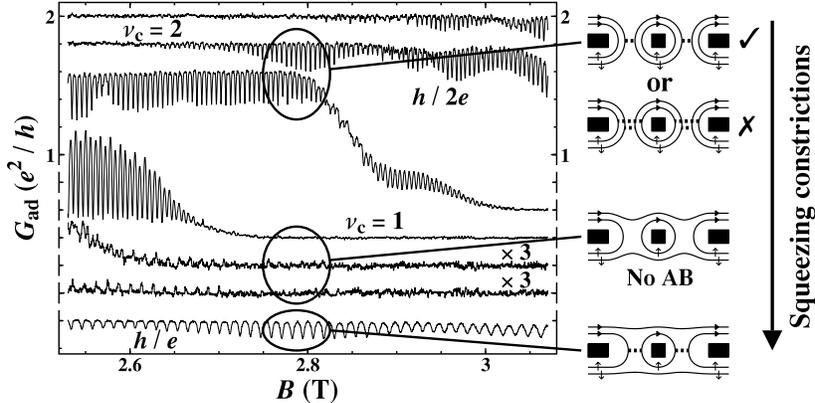}}  

\caption{AB conductance oscillations: the two constrictions were squeezed
symmetrically between traces, which are offset by $0.2e^{2}/h$ down the page
for clarity. Around $B=2.6$~T the pure $h/2e$ oscillations are not completely
established. The diagrams at right show the geometry of edge channels (solid
lines) at around $B=2.8$~T. The black boxes indicate surface gates.
Tunnelling between edge channels is represented by a dotted line.}

\label{fig:DBHIGH}  

\end{figure} 

\end{FIG3}

\end{document}